# Towards MLOps: A DevOps Tools Recommender System for Machine Learning Systems.

Pir Sami Ullah Shah
Department of Software Engineering
*National University of Computer and Emerging Sciences.*
*Islamabad, Pakistan..*

Naveed Ahmad
Department of Software Engineering
*National University of Computer and Emerging Sciences.*
*Islamabad, Pakistan..*

Mirza Omer Beg
Department of Artificial Intelligence
*National University of Computer and Emerging Sciences.*
*Islamabad, Pakistan.*

*Abstract*— Applying DevOps practices to machine learning system is termed as MLOps and machine learning systems evolve on new data unlike traditional systems on requirements. The objective of MLOps is to establish a connection between different open-source tools to construct a pipeline that can automatically perform steps to construct a dataset, train the machine learning model and deploy the model to the production as well as store different versions of model and dataset. Benefits of MLOps is to make sure the fast delivery of the new trained models to the production to have accurate results. Furthermore, MLOps practice impacts the overall quality of the software products and is completely dependent on open-source tools and selection of relevant open-source tools is considered as challenged while a generalized method to select an appropriate open-source tools is desirable. In this paper, we present a framework for recommendation system that processes the contextual information (e.g., nature of data, type of the data) of the machine learning project and recommends a relevant toolchain (tech-stack) for the operationalization of machine learning systems. To check the applicability of the proposed framework, four different approaches i.e., rule-based, random forest, decision trees and k-nearest neighbors were investigated where precision, recall and f-score is measured, the random forest out classed other approaches with highest f-score value of 0.66.

Keywords—Recommender System, MLOps, DevOps, MLOps pipeline, DevOps Tools.

## I. Introduction

Applying DevOps practice to machine learning systems is termed as MLOps. Machine Learning Operations (MLOps) is the combination of practices, tools, and techniques that help organizations manage and deploy machine learning models into production environments. It aims to bridge the gap between data scientists and IT professionals, making it easier to deploy and maintain machine learning models in production. The origins of MLOps can be traced back to the early days of machine learning, when data scientists were responsible for creating and deploying machine learning models. However, as machine learning technology has become more prevalent, the need for a more streamlined and efficient approach to deploying models has become apparent. MLOps emerged as a response to this need, focusing on the integration of machine learning models into production environments and the management of these models over time. It involves the use of automation, continuous integration and delivery, and monitoring to ensure that machine learning models are reliable and efficient.

However, one can face multiple challenges in the construction of MLOps pipeline to deploy machine learning models into production environments. The complete MLOps process and its implementation; is described [1]. MLOps pipeline can be constructed by the combination of open-source tools and the availability of open-source tools is abundant [2] and the number of the tools is expected to be increased further. However, selecting a relevant tech-stack for the construction of DevOps pipeline; is considered as challenge [2]–[6] and sometimes becomes cumbersome while the quality of software products also depends on the DevOps pipeline [6] and it is still desirable to have a generalized guidelines to select appropriate open source tools.

This paper presents a recommender system that recommends open-source tools to construct MLOps pipeline that can automatically train, test and deploy the machine learning models. The proposed system does recommendations on the basis of user inputs i.e., type of data and nature of data. Moreover, the output includes complete details of the recommended MLOps tools that includes the different MLOps phases where the recommend tools can be employed. Furthermore, there three major contributions of the in-hand paper. The first is to construct a dataset of machine learning projects that comprises contextual information i.e., data nature, data type and nature of project which can be utilized in other areas of the research. The second is to develop a recommender system that recommends the MLOps tools based on contextual information of machine learning projects while the last contribution is to evaluate the proposed recommender system in order to know how applicability it is.

This paper is structured into multiple sections where section II describes the complete literature review of the work. Section III describes the methodology; Section IV describes the results and evaluations of the proposed framework. Last but not the least, conclusion is given in section V.

## II. Literature review

Essentially, MLOps is a new phenomenon with very limited literature. Therefore, this section is classified into three different sections as described in Figure 1: MLOps proposed frameworks, applications of MLOps and challenges of MLOps.

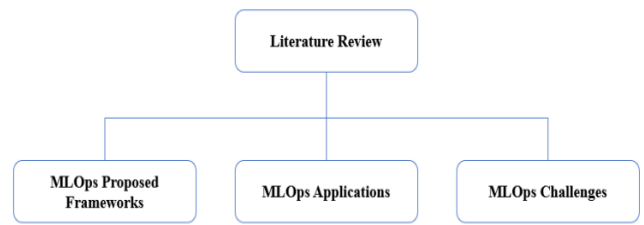

FIGURE 1 LITERATURE REVIEW CLASSIFICATION

*A.    MLOps Proposed Frameworks*

As mentioned earlier that the phenomenon of MLOps practice is very new and it is flourishing as single day passes. Therefore, multiple DevOps frameworks for continuous integrations/deployment of machine learning (ML) and internet of things (IoT) system are proposed. For example, Zhou et al. examined challenges associated in the development and deployment of ML systems and come up with a platform that provides integration of multiple CI/CD tools with Kubeflow to construct a MLOps pipeline and evaluate the performance of the constructed pipeline through multiple controlled experiments [1]. Meenu et al. [7] derive a framework after performing an extensive SLR and GLR review regarding the adoption of MLOps practice, validated the proposed framework by mapping it with three embedded systems case companies [8]. Yan et al. [9] built a platform called OCDL that can be used for constructing, training, reviewing, inspecting, and automatic deployment of machine learning model. Moreover, the design of the proposed platform is completely driven by rules involved in the development lifecycle of machine learning systems. Similarly, S. Makinen [10] has investigated the issues in proprietary software applications that provides services of deploying machine learning applications. The author has presented an open-source cloud native MLOps pipeline that is claimed to be fit to the teams who aim to automate the deployment of machine learning systems. Alonso et al. [11]have investigated issues that can be faced in the operations of multi-cloud applications and proposed a framework called DECIDE that helps the deployment process of multi-cloud applications easier. Moreover, there is a clear surge in applications of artificial intelligence of things (AIoT) devices. Raj [12] has proposed a framework that helps in applying CI/CD practices to machine learning systems right at the edge of AIoT applications and the work has been extended in [13] by performing an extensive experimentation of the proposed framework by checking the air quality of the room for almost 45 consecutive days. The author concluded the obtained results as promising because the proposed framework executed MLOps pipeline without any failure. Similarly, in [14], authors have investigated the continuous monitoring of IoT (Internet of Things) based systems and performs an activity that continuously monitors the devices at production and generates fast feedback to the development in case any errors occur.

*B.    MLOps Applications*

This category of literature delves into practical applications of machine learning operations, illustrating how stakeholders can derive benefits from embracing MLOps in their operational strategies. Sasu et al [15] carried out a study to learn about tech companies worldwide that provide machine learning services. The survey showed that MLOps is crucial in data science, emphasizing on the training of models in production for better and updated results. The process of machine learning operations is very complex in general as it involves automation phases of data engineering including its versioning and its maintenance in remote storages. The overall procedure of machine learning operations is complex, encompassing automated stages of data engineering such as versioning and upkeep in remote storage. Furthermore, it involves the automated experimentation of machine learning models, covering versioning, tracking, comparison, storing of models, and orchestrating pipelines within cloud systems.

Saeed et al. [16] pointed out how complex machine learning operations can be and how it affects the overall quality of the system. They discussed how the continuous integration and continuous delivery (CI/CD) process can be used in deploying machine learning systems. This approach helps in cutting down on development and deployment time while also bringing transparency to the process. Monitoring machine learning models in production is considered challenging because of high cost and large training time. Silva et al. [17] proposed a framework that provides different guidelines which help in effective monitoring of machine learning models in production. The proposed approach is evaluated on 9 different applications which concluded the clear visibility of the work.

*C.    MLOps Challenges*

This category of literature outlines the challenges encountered in the development and deployment of machine learning systems. Most of these challenges are associated with the in-availability of the skill set and the widespread availability of open-source tools. Ref et al. [18] explored various roles and tools essential for building an MLOps pipeline. Additionally, they conducted a comparison of 26 different tools, considering the phases of the MLOps pipeline and evaluating certain quality attributes such as usability. Moreover, Gill et al. [19] conducted a systematic literature review, systematically identifying, reviewing, and synthesizing studies published in the public domain between 2010 and 2016. The authors found numerous advantages in adopting DevOps practices. However, they also highlighted significant challenges, particularly in the selection of suitable tools for constructing pipelines. This tool selection process can have a direct impact on the overall product quality, thereby influencing organizational values. Most of time multiple cloud services are utilized to deploy the different parts of the overall applications however the consideration of different cloud services depends on multiple factors i.e., less services charges, latency time, easy to operate and much more. Similarly, machine learning systems can be deployed to production by considering multiple cloud services which is called as multi-cloud environments which is considered difficult and time consuming as claimed by Banerjee et al. in [20]. The performance of machine learning systems relies on MLOps as mentioned in [18]. In a similar vein, Tamburri [21] has recognized the significance of MLOps in shaping the quality attributes and has outlined the challenges associated with enhancing the sustainability of AI-enabled systems through MLOps practices. In the current landscape, it has become a common practice to employ hybrid machine learning models, where multiple models collaborate for a single purpose. This collaborative approach often involves the participation of various organizations. In alignment with this trend, Granlund et al. [22] have specifically addressed challenges related to implementing MLOps for AI-based systems developed by multiple organizations. Notably, these challenges center around integrating diverse models from different organizations to address a singular problem. Additionally, the authors have proposed standard patterns that could prove valuable in facilitating the smooth deployment of such systems across a multi-organizational context. Hewage and Meedeniya [23] delved into the technical obstacles to the development and delivery of machine learning systems, recommended MLOps as a best practice to cope with all the identified problems.

## III. STUDY METHODOLOGY

### A. Construction of Dataset

**Dataset Collection.** We collected data in two different phases as in Figure 2. Initially, data of final year projects was collected from students of the final year in the campus. The collection of data was done through online google form, detail to each question included in the form is described in TABLE I. Furthermore, data of more machine learning projects was collected from online repositories i.e., GitHub and Kaggle. The total number of collected projects both artificial intelligence (AI) and non-artificial intelligence based was 145.

TABLE I. DATA COLLECTION FORM

| Q# | Description |
|---|---|
| Q1. | Name of the Project |
| Q2. | Description of the Project |
| Q3. | Type of the Project |
| Q4. | Nature of data |
| Q5. | Type of data |
| Q6. | Preprocessing Tools used to process the data. |
| Q7. | Type of the Project |
| Q8. | ML/ DL (Technique) |
| Q9. | Evaluation Metrics |
| Q10. | Tools used to construct model. |

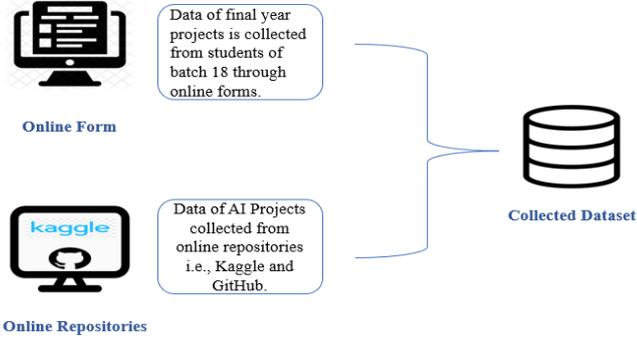

FIGURE II DATA COLLECTION PROCESS

**Dataset Preparation.** To separate AI projects from non-AI projects, filtration is performed where 108 projects were found as AI based projects while the remaining were found as non-AI projects. Moreover, features of dataset were 10 in total and two of them were selected as input features i.e., type of data and nature of data while four of them were selected as output features i.e., preprocessing tools, model construction tools, type of project and evaluation metrics. Preparation of dataset is visually represented in Figure III.

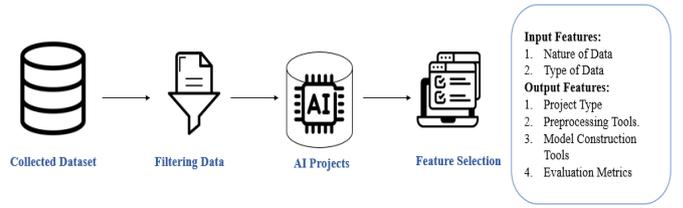

FIGURE III DATA PREPARATION PROCESS

**Dataset Processing.** Before constructing machine learning models, the data undergoes preprocessing, which involves two main steps as visually described in Figure IV: vectorization and data partitioning into training and testing sets. Initially, vectorization is carried out using the one-hot encoding technique. Subsequently, the dataset is split into two parts with an 8:10 and 2:10 ratio. The larger portion, constituting 8:10 of the data, is designated as the training data, serving as the basis for training the machine learning models. The smaller segment, comprising 2:10 of the data, is referred to as the validation or test data. This set is employed to assess the performance of the trained models and ensure their ability to generalize to new data.

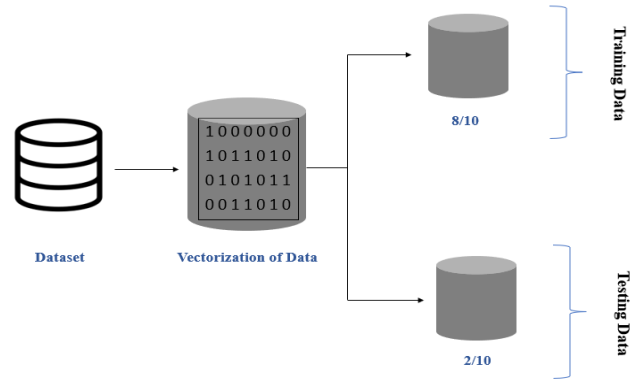

FIGURE IV DATASET PROCESSING

### B. Recommender System Development Phases

Figure V describes all the phases that were performed to develop recommender system. After the dataset creation and preprocessing, the subsequent step involved the extraction of rules from the data, resulting in a total of 24 different rules. Moreover, multiple approaches were employed—three model-based and one rule-based. Figure VI provides a comprehensive breakdown of the extracted rules. Each rule, grounded in the nature and type of data, produces four outputs: evaluation metric, preprocessing tools, model construction tools, and the type of project. In the construction of machine learning models, as highlighted earlier, the data underwent vectorization and was partitioned into training and testing sets, with an 80% and 20% ratio, respectively. Three distinct machine learning models—random forest, decision trees, and K-Nearest Neighbors—were then trained using the designated training data. Subsequently, each approach underwent evaluation on the testing data. The evaluation of each approach involved the measurement of three key metrics: recall, precision, and f-score. These metrics provided a comprehensive understanding of the models' performance. Notably, the selection of the best approach was based on the f-score, with the random forest model achieving the highest value among all evaluated models. This thorough approach to model evaluation ensures a robust and informed selection process for the most effective machine learning approach in the context of the extracted rules. Lastly, the endpoint to the model with highest f-score was created and integrated with modules of the recommender system.

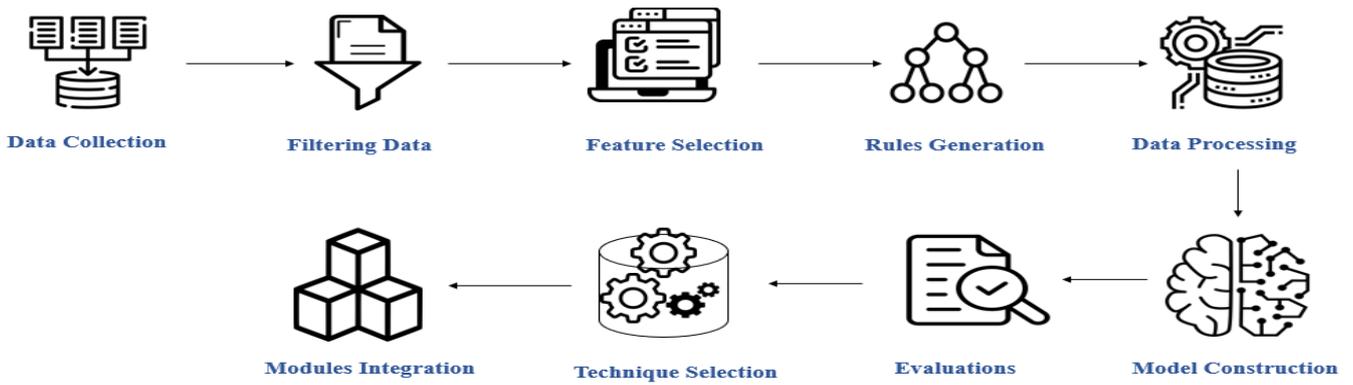

FIGURE V DEVELOPMENT PHASES

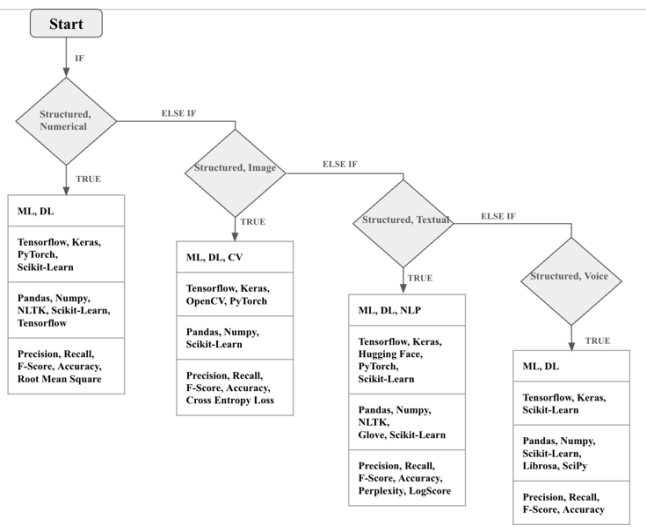

FIGURE VI EXTRACTED RULES

### C. Recommender System Overview

The proposed recommender system offers recommendations for open-source tools to operationalize machine learning systems based on contextual information. Initially, this context includes the type of data (numerical, textual, image, or video) and the nature of data (structured, unstructured, or semi-structured). The system predicts preprocessing and model construction tools based on this contextual information. As mentioned earlier, the technology stack of open-source tools for constructing MLOps pipelines is vast. Each specific tool integrates with various third-party tools, such as ZenML, which connects with scikit-learn, Keras, TensorFlow, and more.

The proposed framework maps the predicted preprocessing and model construction tools with MLOps tools available in the catalogue, recommending matching tools to the machine learning engineer. Figure VII provides a visual representation of this recommender system.

Consider a machine learning engineer from Company X who aims to construct an MLOps pipeline for automating the development process of a machine learning system. However, due to the abundance of open-source tools, they struggle to select a relevant chain of tools. Through the proposed recommender system, the machine learning engineer inputs contextual information, specifying the type and nature of the data for the project. Initially, the system predicts tools for data preprocessing and model construction. Subsequently, the MLOps tools are investigated, involving a step to verify their integration with the tools predicted by the model. Upon successful integration with the predicted preprocessing and model construction tools, the recommended MLOps tools are suggested to the machine learning engineer for seamless implementation.

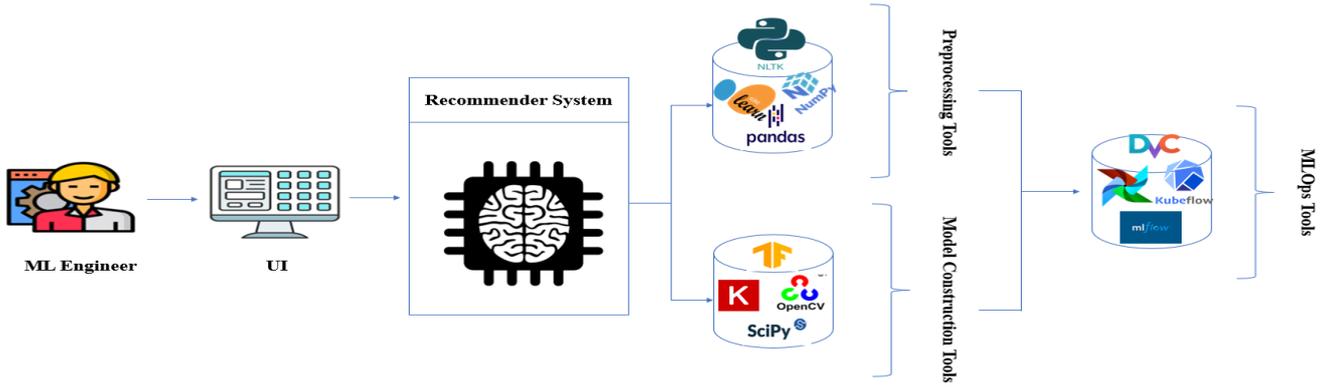

FIGURE VII RECOMMENDER SYSYEM OVERVIEW

*D. Results and Evaluation*

As previously mentioned, we explored four different approaches to address the given problem. These approaches underwent a comprehensive comparison and evaluation using a designated set of testing data. The output labels of the testing data were aligned with the values obtained from each approach. Consequently, we measured three distinct evaluation metrics for each approach, namely precision (1), recall (2), and f-score (3), as outlined in TABLE II.

Precision gauges the proportion of true positive results out of all positive results, recall measures the proportion of true positive results out of all actual positive instances, and the f-score represents a harmonic mean of precision and recall. By selecting the approach that yielded the highest f-score, we opted for the utilization of the random forest algorithm, integrating it into the final product. The f-score approach provides the best balance between precision and recall, making it the most suitable for addressing the specific problem at hand.

In Figure VIII, the diagram illustrates three distinct bars for each approach—rule-based, random forest, decision forest, and K-Nearest Neighbors—displayed along the x-axis. This visual representation further clarifies the comparative performance of each approach in terms of the specified evaluation metrics.

$$Precision = \frac{TP}{TP + FP} \quad (1)$$

$$Recall = \frac{TP}{TP + FN} \quad (2)$$

$$F - measure = \frac{2 * precision * recall}{precision + recall} \quad (3)$$

Where,

- TP stands for True Positives, are the instances that are correctly classified as positive. In other words, correct prediction of positive class.
- FP stands for False Positives, are the instances that are incorrectly classified as positive. In other words, a wrong prediction of positive class.
- FN stands for False Negatives, are the instances that are incorrectly classified as negative. In other words, wrong prediction of negative class.

TABLE II. EXPERIMENT RESULTS

| Approach | Precision | Recall | F-Measure |
|---|---|---|---|
| Decision Tress | 0.683871 | 0.582418 | 0.629080 |
| Random Forest | 0.705521 | 0.631868 | 0.666667 |
| K-Nearest Neighbors | 0.674556 | 0.626374 | 0.649573 |
| Rule-Based | 0.498423 | 0.868132 | 0.633267 |

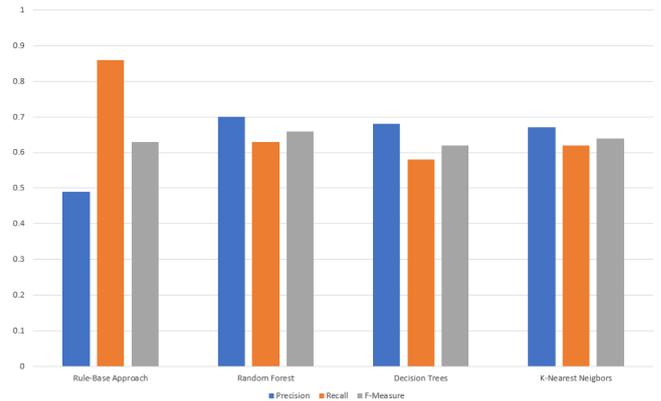

FIGURE VIII PRECISION, RECALL, F-MEASURE

## IV. CONCLUSION

In order to fully automate the development process of machine learning systems, a practice is used known as MLOps. Through literature, multiple challenges are identified that hesitate machine learning engineers to adopts the practice. Although it impacts the overall development of the systems and the quality of the concerned machine learning project. One of the major challenges, is to select a relevant tech stack in order to construct MLOps pipeline in other words the availability of open-source tools is abundant and it is claimed to be further increased. To cope this problem, a framework for recommender system is proposed that recommends open-source tools that can help to construct a MLOps pipeline through which the organizations can provides instant deliveries of new features with high quality values. However, the proposed framework works as collaborative filtering (recommendation engine) that recommends on the basis of other similar users. To evaluate the proposed framework, experimentation is performed on testing data in order to measure recall, precision and f-score. Moreover, the values of evaluation metrics can be improved by increasing the number of records in the dataset. Importantly, the second contribution of the in-hand research is the construction of a dataset that contains the data in other words the contextual information of machine learning projects. The future direction of the given research is to investigate the quality of the machine learning system on the basis of MLOps pipeline.